\documentstyle[prl,aps,epsfig,floats]{revtex}
 
\title {Low energy magnetic excitations of the Mn$_{12}$-acetate spin cluster observed by neutron scattering } 
\author{I. Mirebeau$^1$, M. Hennion$^1$, H. Casalta$^2$, H. Andres$^3$, H. U. G\"udel$^3$
 A. V. Irodova$^4$, and A. Caneschi$^5$}

\address{1 - Laboratoire L\'eon Brillouin, CEA-CNRS, CE Saclay, 91191 Gif sur Yvette, France}
\address{2 - Institut Laue Langevin BP 156 F-38042 Grenoble, France}
\address{3-Department of Chemistry, University of Bern, 3000 Bern 9, Switzerland}
\address{4 -Russian Research Center, Kurchatov Institute 123182 Moscow, Russian Federation}
\address{5 -Department of Chemistry, University of Florence, Via Maragliano 7, 50144, 
Firenze, Italy}

\date{\today, submitted to Phy. Rev. Lett.}
\begin{document}
\twocolumn[\hsize\textwidth\columnwidth\hsize\csname @twocolumnfalse\endcsname
\maketitle
\begin{abstract}
We performed high resolution diffraction and inelastic neutron scattering measurements of Mn$_{12}$-acetate. Using a very high energy resolution, we could separate the 
energy levels corresponding to the splitting of the lowest S multiplet. Data were analyzed 
within a single spin model (S=10 ground state), using a spin Hamiltonian with parameters up to 4$^{th}$ order.
 The non regular spacing
 of the transition energies unambiguously shows the presence of high order terms in the anisotropy
 (D= -0.457(2) cm$^{-1}$, B$_4^0$= -2.33(4) 10$^{-5}$cm$^{-1}$). 
 The  relative intensity of the  lowest energy peaks is
 very sensitive to the small transverse term, supposed to be mainly responsible for quantum tunneling.
   This allows an accurate 
determination of this term in zero magnetic field (B$_4^4$=$\pm$3.0(5) 10$^{-5}$
cm$^{-1}$). The neutron results are discussed in view of
 recent experiments and theories.   
\end{abstract}
]

 Studies of molecular nanomagnets are rapidlly growing. The most
attractive clusters consist of a few (typically 10-20) paramagnetic ions
coupled by exchange interactions, therefore at the borderline between quantum
and classical behavior. Besides their great theoretical interest, the study
of these magnetic molecules could have important practical consequences, since it
helps to determine the size limit for information storage.
Mn$_{12}$-acetate is the best studied spin cluster so far. In a simple 
 ionic picture, an external ring
of eight Mn III ions (S$_2$=2) surrounds a tetrahedron of four Mn IV ions
(S$_1$=3/2), giving the molecule a plate-like shape with tetragonal
symmetry\cite{Lis}. Above 10K, a reversal of the molecular
magnetization occurs by thermally overcoming the energy barrier due
to uniaxial anisotropy (superparamagnetic behavior) \cite{Sessoli}.
 Below 2K, this reversal is governed by a quantum tunneling 
mechanism as shown by a finite relaxation time in the magnetization at T=0\cite{Hernandez}. Mn$_{12}$-ac also exhibits as a spectacular effect, regular steps in the
hysteresis cycle \cite{Thomas}. These observations have been interpreted by
tunneling processes within the ground state of the molecule \cite{Politi,Fort,Chudnovsky,Gunther}.
 As a peculiarity of this cluster, the tunneling process is thermally
assisted. Mn$_{12}$-ac is usually described as a single spin with
S=10 ground state, split by anisotropy terms into sublevels with (-10$\leq$M$\leq$10).
This picture relies on the assumption that the anisotropy energy is
significantly smaller than the exchange energy. The S=10 ground state
corresponds to a ferrimagnetic spin arrangement of the two types of  Mn 
 ions \cite{Sessoli}, \cite {Pederson99}.
On this basis, high field-high frequency EPR measurements were
analyzed  by considering spin Hamiltonian parameters up to 4$^{th}$ order\cite{Barra97}. This gave the first experimental
evidence of the transverse term searched for to explain the  quantum tunneling.  However, depending on the range of frequency or applied field
considered, different sets of parameters were reported \cite{Barra97,Hills98}. Very recently, the position of the first three excited sublevels of the S=10 ground state was determined by optical spectroscopy in the far infrared (IR) range \cite{Mukhin98}.

 In this work, we present a detailed study of the energy sublevels of the ground state by inelastic neutron scattering (INS). Neutron scattering is the most powerful tool to investigate spin structure
and spin excitations. In contrast to EPR, it is a zero field experiment, which requires no assumption about
the Land\'e factor g.  INS has been used to determine exchange and anisotropy splittings in numerous clusters of transition and rare earth metal ions \cite{Gudel96}. Very recently, it was used to determine the zero field splitting in a Fe$_{8}$ cluster \cite{Cac98}. In Mn$_{12}$-ac, we previously studied
spin excitations up to 12 $meV$ \cite{Hennion97}. We observed several excitations attributed
either to the anisotropy or to the exchange terms, and determined their
dynamical form factor. The position of the anisotropy level at low temperature
 (1.24 $meV$)
agreed with the EPR determination, but we were not able
to resolve the detailed structure of the other sublevels at higher temperature.

 We present here results obtained with a very high energy
resolution, which enable us to separate the sublevels of
the ground state. The non regular spacing of the transition energies
unambiguously shows the presence of high order terms in the spin Hamiltonian. We
determine these parameters up to 4$^{th}$ order within the single S=10 ground state model.
Interestingly, the intensities of the lowest energy excitations associated
with levels near the top of the barrier, are very sensitive to
the value of the transverse 4$^{th}$ order term. This allows an accurate
determination of this very important term, although its value is very small.

 A
 [Mn$_{12}$O$_{12}$(CD$_3$COO)$_{16}$(D$_2$O)$_4$].2CD$_3$COOD.4D$_2$O 
 powder sample was synthetized
using deuterated solvents, since hydrogen has a large incoherent cross
section which is the main background in neutron experiments. A deuteriation of 93$\%$ was  achieved, as shown by the analysis of the neutron diffraction data. The crystal
structure was tested between 10K and 290K by neutron diffraction on the
diffractometer G6.1 at the Laboratoire L\'eon Brillouin. Data were analyzed with the FULLPROF program \cite{Rodriguez}.
The structure keeps the tetragonal symmetry
in the whole temperature range.
Above 130 K the framework of the molecule, consisting of Mn, C and O atoms
practically coincides with that determined by single crystal X-ray diffraction\cite{Lis}. Only
slight displacements of the acetic acid and water
molecules of solvation were found. We precisely stated the positions of the H/D atoms,
using common constraints for interatomic distances H-O and H-C, and angles
H-O-H and H-C-H. Moreover, we determined the unknown H positions in the
acetic acid molecules of solvation. Only the water molecules contain
hydrogen (about 33.5$\%$), whereas the rest of the cluster is fully deuterated. Below 130
K, we found a slight distortion of the cluster framework caused by displacements of
the bridging acetate ligands. The stronger distorsion of the octahedral oxo coordination reduces the local symmetry of Mn III
with respect to room temperature, the environment of Mn IV
 being preserved (besides the usual lattice contraction).
 
INS measurements were performed with the high energy resolution
time of flight spectrometer IN5 at ILL, in the temperature range 1.5K-35K.
The incident wavelength  was 5.9$\AA$. The resolution of the experiment at zero
energy transfer  (HWHM=27.5$\mu eV$)was determined using a vanadium standard. 
In the energy range studied (-1.5,1.5) $meV$, we checked that the q
dependence of the neutron cross section is the same for all the inelastic transitions.
It varies very little with temperature in the range (0.2, 2) $\AA$$^{-1}$ and agrees
with previous results obtained by performing constant-q scans on  a triple
axis spectrometer \cite{Hennion97}. We therefore added the contributions of all detectors,
 which strongly increased the statistical accuracy.
 The experimental spectra are shown in Fig.1 for three temperatures. At
T=1.5K, a well defined excitation is observed at 1.24 $meV$ on the sample energy gain side
($\hbar\omega$$>$0), with a peak
width limited by the experimental resolution. It is readily
attributed to an excitation from the lowest energy level (M=$\pm$10) to the
first excited level (M= $\pm$9). The peak at 0.9 $meV$, much less intense than in previous samples \cite{Hennion97}, was not considered in the analysis. We also note a spurious temperature independent signal at the foot of the elastic peak. It was fitted by a gaussian and substracted. As the temperature increases, higher energy levels are populated, so that new excitations appear on both sides of the spectrum. At 23.8K, we distinctly observe 14 well separated
peaks, together with much smaller ones around $\pm$0.2 $meV$ (see the inset of Fig.1).
 The position of the
well resolved peaks is temperature independent. Their width, 
close to the resolution limit, increases with temperature
of about 25$\%$ between 1.5K and 25K, suggesting a finite life time. The energy intervals between adjacent peaks in the spectrum show a small but significant decrease with decreasing energy transfer. Below 25K, the temperature evolution of the peak intensities
reflects the depopulation of the lowest level at the benefit of the excited
levels within the ground state manifold. Above 25K, all peaks broaden and the
total inelastic intensity decreases, as the levels attributed to higher spin states become populated.

\begin{figure}
\centerline{\epsfig{file=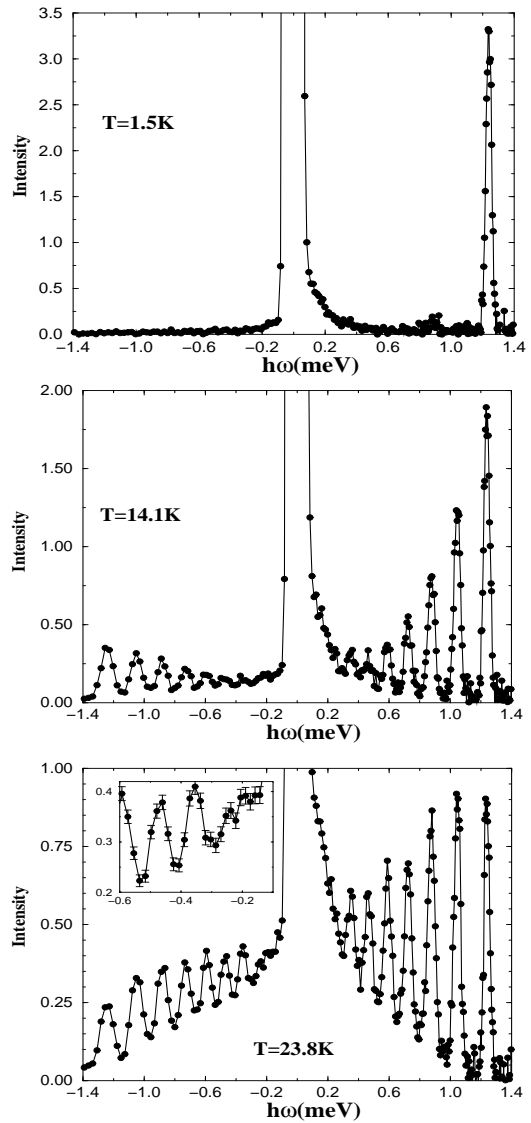,height=15 cm,width=7cm}}
\caption{Energy spectra at 3 temperatures (raw data). The intensity units are the same but the scales are different for all spectra. In inset, the low energy range on the sample energy loss side.}
\label{fig:38_6.8} \end{figure}

In the approximation of a  S=10 single spin Hamiltonian, the magnetic intensity is written as \cite{Birgenau}:\\
\\
 \\
S({\bf Q},$\hbar\omega$)=N($\gamma$$_N$r$_e$/2)$^2$g$^2$f$^2$({\bf Q})$\sum \limits_{i,f}$p$_i$$|$$<$f$|$S$_\perp$$|$i$>$$|^2$$\delta$($\hbar\omega$-(E$_f$-E$_i$)), 
  with $|$$<$f$|$S$_\perp$$|$i$>$$|^2$=1/3(2$|<f|S_z|i>|^2$\\
\\
\centerline{+$|<f|S_+|i>|^2$+$|<f|S_-|i>|^2$) (1)} \\                      
\\
S$_\perp$ is the spin component perpendicular to the scattering vector {\bf Q}. 
  p$_i$=exp(-E$_i$/k$_B$T)/$\sum\limits_i$exp(-E$_i$/k$_B$T) is the Boltzmann factor of the level i.The delta function is convoluted by the experimental resolution function. The eigenstates $|$i$>$ have the energies E$_i$ obtained by a diagonalization of the spin Hamiltonian. We adopt the spin Hamiltonian used to interpret EPR\cite{Barra97}, magnetization\cite{Barbara98}
and far IR experiments \cite{Mukhin98}\\
\\
$ H= D[S_z^2-1/3S(S+1)]+B_4^0O_4^0+B_4^4O_4^4$ (2)\\
\\
with $O$$_4^0$=35S$_z^4$-[30S(S+1)-25]S$_z^2$-6S(S+1)+3S$^2$(S+1)$^2$ and $O$$_4^4$=1/2(S$_+^4$ + S$_-^4$).

 From equation (1), it follows that only the transitions with $\Delta $M=0 or $\pm$1
are allowed. When only the diagonal terms are considered in equation (2) (B$_4^4$=0), the S=10
manifold splits into 11 sublevels corresponding to M=$\pm$10, ±9,.. 0.The $\pm$M degeneracy is not lifted, and when all possible
levels are populated we expect nine $\Delta $M $\pm$1 transitions on either side of the spectrum. Seven of these correspond to the well resolved peaks observed at 15K and 23.8K in Fig. 1. The decreasing energy spacing with decreasing energy transfer arises mainly from
the 4$^{th}$ order B$_4^0$ term.
\begin{figure}
\centerline{\epsfig{file=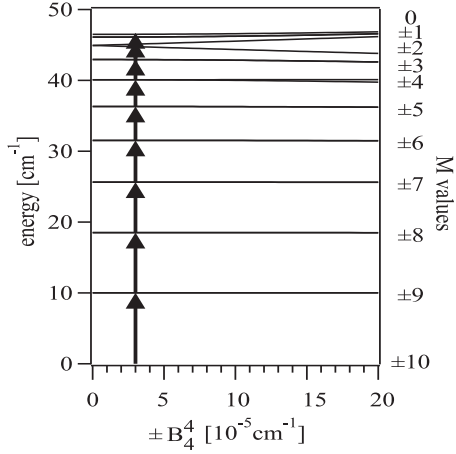,height=6cm,width=6cm}}
\caption{Energy sublevels of the S=10 ground state versus B$_4^4$, as calculated for
D= -0.457cm$^{-1}$,  B$_4^0$= -2.33 10$^{-5}$cm$^{-1}$}
\label{fig:38_6.8} \end{figure}
\begin{figure}
\centerline{\epsfig{file=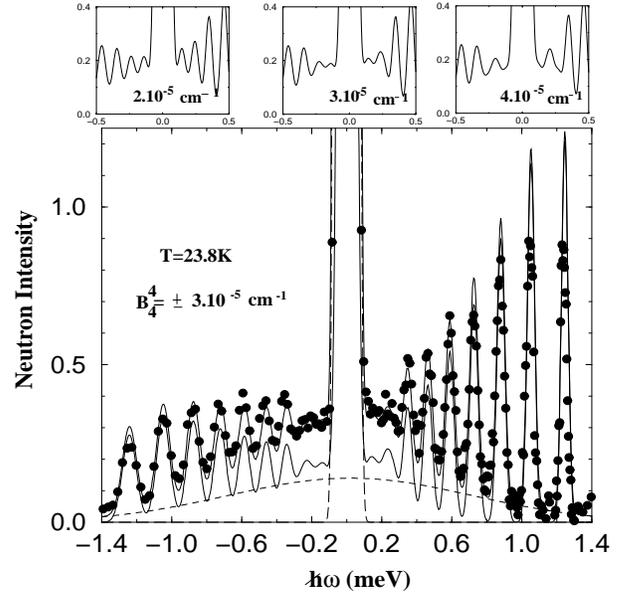,height=8cm,width=8cm}}
\caption{Energy spectra: filled circles are data corrected for the gaussian signal observed at 1.5K near zero energy transfer. The intensity calculation (thin line), lorentzian background (dashed line), elastic intensity (long-dashed line), and the sum of all components (thick line) are also plotted. At the top of the figure, intensity calculations for 3 B$_4^4$ values.}
\label{fig:38_6.8} \end{figure}

 As a first step in our quantitative  data analysis in terms of eqs (1, 2),
we neglected the transverse term B$_4^4$, 
known to be small. The positions  and relative intensities of the five most
intense excitations are well reproduced, and the fitted values of the
parameters (D= -0.457(2) cm$^{-1}$, B$_4^0$= -2.33(4) 10$^{-5}$cm$^{-1}$) 
agree with the determination of Barra et al by
 EPR\cite{Barra97} and Mukhin et al by far IR spectroscopy \cite{Mukhin98}.
 However, this simplified model overestimates the intensities of the excitations below 0.3 $meV$, 
which involve  energy levels close to the top of the
anisotropy barrier.

 In a second step we therefore took the off-diagonal term
B$_4^4$ into account. A non zero B$_4^4$ induces a mixing of energy levels with
different M that becomes more and more effective on approaching the top of
the barrier. This is illustrated in Fig 2 where the positions of the
energy levels E$_i$ are plotted versus B$_4^4$, for given B$_4^0$ and D.
The energies of the highest excited states strongly depend on B$_4^4$ and the
degeneracy of the M=$\pm$2 and $\pm$4 levels is lifted for non-zero B$_4^4$. At the top of Fig 3, we show the
calculated intensities at 23.8K for three different B$_4^4$ values, using the experimental
 energy resolution, in the energy transfer range (-0.5, 0.5) $meV$. In this range the position, shape, and intensities of the inelastic peaks strongly depend on B$_4^4$, whereas
the other peaks remain unaffected. Depending on the B$_4^4$ value (3 or 4 10$^{-5}$cm$^{-1}$), the energy transfer at 0.2 $meV$ corresponds either to a minimum or to a maximum intensity, respectively. Therefore, only B$_4^4$ values in a very narrow range
can explain the experimental behavior. We find that B$_4^4$=$\pm$ 3.0(5) 10$^{-5}$cm$^{-1}$ 
gives the best
agreement with the data. This value is slightly lower and more accurate than the one reported by 
Barra et al (B$_4^4$=$\pm$4(1) 10$^{-5}$cm$^{-1}$). The sign of B$_4^4$ is undetermined, 
owing to the symmetry of the Hamiltonian. In Fig 3, we present the results of the model calculation
with the best fitted values D= -0.457cm$^{-1}$, B$_4^0$= -2.33 10$^{-5}$cm$^{-1}$ and B$_4^4$= $\pm$3.0(5) 10$^{-5}$cm$^{-1}$. In order to compare with experimental data at 23.8K, we added a  lorentzian quasielastic background, probably due to hydrogen. The agreement with experiment is excellent. We note a small reduction of the observed intensities for the excitations near 1.2 $meV$. 
It is ascribed to a partial population of higher energy cluster spin states. The single spin S=10 model breaks down at higher temperatures \cite{Mukhin98}. 
  
 The main result coming out of our analysis is the precise determination in zero field
of the coefficient B$_4^4$ in the transverse term. The presence of non-diagonal terms 
in the spin Hamiltonian
has been searched for by many theoreticians  \cite{Politi,Fort,Chudnovsky}
and experimentalists, since only components which do
not commute with S$_z$could induce tunneling. In zero field,
a term C(S$_+^4$ + S$_-^4$) is the lowest order spin Hamiltonian allowed by
tetragonal  symmetry. Such a term allows tunneling with $\Delta$M=$\pm$4,  as observed
experimentally in the magnetization. Thus, our determination of an accurate  B$_4^4$ value
 in the absence of a magnetic field has a direct relevance for a quantitative understanding
 of the dominant tunneling process. However, other transverse terms are required to explain the
tunneling transitions with $\Delta$M=$\pm$1. Politi et al.
 \cite{Politi} have also
argued that if only this 4$^{th}$ order transverse term was involved,
 a very small field like the earth field
 would destroy tunnelling. Other
transverse terms deriving from dipolar coupling \cite{Luis98}, 
hyperfine interactions \cite{Barbara98}, 
\cite{Luis98} 
or a spin-phonon coupling mechanism
 \cite{Politi}, \cite{Fort} have been suggested in addition to the previous one.
 Up to now, they have not been observed experimentally, 
and our results provide no evidence for their existence.

The analysis presented above is based on the assumption of a "single
spin" ground state.  This simple model describes the low energy
excitations presented in this work very well. However, our earlier INS
experiments at higher energies \cite{Hennion97} show that excitations involving other spin states
are rather close to the S$=$10 ground state. Recent measurements by magnetization \cite{Barbara98},  or heat capacity \cite{Gomes98} also point out the influence of excited states with
different S values.
 Several microscopic descriptions of the S=10 ground
state have been proposed \cite{Sessoli}, \cite{Barra97}, \cite{Zvezdin}, which
 involve different coupling schemes between the individual spins. Due to the large
number of cluster spin states, 10$^{8}$ in the general
case, simplifications must be made by considering the hierarchy of the
exchange interactions between the individual spins. Very recently, a new formalism for the energy
has been proposed \cite{Katsnelson99}. Following the "Florentine" coupling scheme \cite{Sessoli}, the Mn$_{12}$ cluster is described by four dimers
 Mn$^{3+}$-Mn$^{4+}$
with spin s=1/2, and four Mn$^{3+}$  ions (S=2), coupled by isotropic exchange
interactions. The dominant exchange coupling is between the spins within each
dimer, as first suggested by \cite{Sessoli}, and in agreement with recent magnetic
measurements in the MegaGauss range \cite{Mukhinunp}. The Hamiltonian includes relativistic anisotropic interactions, the Dzyaloshinskii-Moriya ones being the most important. This model qualitatively explains the spin excitations previously measured by INS up to 12 $meV$ \cite{Hennion97}.  We believe that our
new experimental results will be a crucial checkpoint for further theoretical work.

\end{document}